\documentclass[aps,prl,twocolumn,showpacs]{revtex4}
\usepackage{amsmath}
\usepackage{amssymb}
\usepackage{graphicx}
\usepackage{dcolumn}
\usepackage{bm}

\begin{document}

\title{Reply  to the Comment, arXiv:0810.3243v1 by 
B.~Geyer, G.~L.~Klimchitskaya, U.~Mohideen, and V.~M.~Mostepanenko}
\author{L.~P.~Pitaevskii}

\begin{abstract}
It is shown that criticism of paper \cite{Pitaevskii} by the authors of the Comment \cite{Comment} is wrong and 
and that their main
arguments are in  contradiction with 
established concepts of statistical physics.
\end{abstract}

\pacs{34.35.+a,42.50.Nn,12.20.-m}
\affiliation{CNR INFM-BEC, Department of Physics, University of Trento, I-38100 Povo,
Trento, Italy;\\
Kapitza Institute for Physical Problems, Kosygina 2, 119334 Moscow, Russia.}
\maketitle

1. The main point of the Comment is that the theory developed in \cite{Pitaevskii}
''leads to the violation of Nernst's theorem'' for
 materials where ''the
concentration of charge carriers, $n$, does not go to zero when temperature vanishes, but the
conductivity  goes to zero due to vanishing mobility $\mu$''.

I believe that this statement is a
result of a pure misunderstanding. The materials under discussion are
amorphous glass-like \textit{disordered} bodies. Conductivity goes to zero
with $T$ in such materials due to localization of the charge carriers just
because of the disorder. The point is that Nernst's theorem \textit{is not
valid} for these disordered bodies. It is well-known that they have a big finite entropy at zero
temperature. Localized carriers also contribute to this residual entropy and
the calculation of a small correction to its value due to the Casimir-Polder
interaction scarcely has a physical meaning. Of course, the existence of
disordered bodies at $T=0$ itself does not contradict quantum statistical mechanics. They
are simply not at an equilibrium state at low temperatures due to a very
long relaxation time. Particularly they are not in their ground state at $T=0$.
The criticism of an application of the original Lifshitz theory to these disordered
bodies on the  basis of 
Nernst's theorem   in 
\cite{KG08} is also wrong for the same reason.

2. The authors state: ''Physically, the theory of \cite{Pitaevskii} includes the
effect of screening, i.e., nonzero gradients of $n$. This situation is out of thermal equilibrium
which is the basic applicability condition of the formalism of \cite{Pitaevskii}''.

This statement is obviously wrong and odd. It is
well-known that the Boltzmann distribution in the electric field, which was
used in \cite{Pitaevskii} for describing screening, is an {\it equilibrium}
distribution. Actually, in equilibrium, the diffusion current 
due to the gradient of $n$ is compensated
by the mobility of carriers due to an electric field. 
This compensation results in Einstein's relation between the
coefficients of diffusion and mobility.

3. The authors say that papers \cite{Pitaevskii,Lam,Svetovoy} violate
''the Nernst
theorem for metals with perfect crystal lattices''.
This severe statement is based, however, on the paper \cite{Bez2}, which is
certainly wrong, 
because the authors used the {\it normal} skin effect theory for metals
with perfect crystal lattices at $T \to 0$, while it is well-known that in this situation
one must use the {\it anomalous} skin effect theory (see
\cite{Kit}). 
Consequently it is impossible to make any conclusions on the basis of this paper.

4. The authors object to my statement  that ''it is difficult to estimate the
number of ions which are effective in mobility and screening'' on the basis that $n$
''can be obtained by the method'' presented in \cite{tomozawa}. However, the results of the 
paper \cite{tomozawa} confirm my point of view. At given number of impurities atoms 
$n_{Na}\sim 2 \times 10^{15}$cm$^{-3}$ and at $T=433$K, 
the number of carriers $n$  in fused silica changes from  $3 \times 10^{12}$cm$^{-3}$
to $2 \times 10^{8}$cm$^{-3}$ depending of the water content. Furthermore, the 
measured temperature dependence of $n$ is in strong contradiction with the authors'
assumptions. When temperature decreases from 473K to 433K, $n$ {\it decreases} from 
$6 \times 10^{13}$cm$^{-3}$  to $3 \times 10^{12}$cm$^{-3}$ and the authors of \cite{tomozawa} explicitly
 conclude: ''The change in conductivity nearly scales with the change
of charge carrier concentration suggesting that {\it the mobility remains nearly independent
of temperature}''\cite{footnote}.

5. I do not have enough information about details of the experiment and calculations \cite{chen} to discuss their accuracy. 
In general, if an experiment and a theory do not agree, it does not always mean that the theory is wrong. 

 The authors  believe that my statement that for the relaxation time of the order of 917 hours
''the carriers mobility can hardly be important in any experiments'' agrees with their prescription ''that 
for dielectrics the dc conductivity should be disregarded''. However authors of \cite{chen} disregard the dc conductivity
only in absence of light and take it into account in the presence of light, without any foundations for this difference.
In my opinion this procedure is not consistent.
In fact, I wanted only to stress that the screening plays no role if the relaxation
time is larger than the duration of an experiment. 
However, I see no reason not to take into account the conductivity if equilibrium is reached. 

In conclusion I would like to emphasize that I do not think that all problems of the thermal Van der Waals forces
are clear both from  the theoretical and experimental points of view. However, their solution cannot be based on discarding established concepts of statistical physics.

I thank M.~Tomosawa for useful discussion properties of fused silica.


\begin{thebibliography}{99}
\bibitem{footnote} I do not know if it is possible to extrapolate this dependence to $T=0$. As it follows from
item 1, it makes no difference for our discussion.
\end{thebibliography}
\end{document}